\documentclass{amsart} 
\usepackage{amssymb,amsmath,tikz,mathtools}
\DeclareMathOperator{\Spec}{Spec}
\DeclareMathOperator{\Proj}{Proj}
\DeclareMathOperator{\AProj}{AProj}
\DeclareMathOperator{\Aut}{Aut}
\DeclareMathOperator{\Sym}{Sym}

\newcommand{\cC}{\ensuremath{\mathcal{C}}}
\newcommand{\cV}{\ensuremath{\mathcal{V}}}
\newcommand{\R}{\mathbb{R}}
\newcommand{\C}{\mathbb{C}}
\newcommand{\M}{\mathbb{M}}
\newcommand{\op}{\ensuremath{^{\mathrm{op}}}}
\newcommand{\commeas}{\ensuremath{\odot}}

\numberwithin{equation}{section}
\theoremstyle{plain}
\newtheorem{Theorem}[equation]{Theorem}
\newtheorem{Lemma}[equation]{Lemma}
\newtheorem{Proposition}[equation]{Proposition}
\newtheorem{Corollary}[equation]{Corollary}
\theoremstyle{definition}
\newtheorem{Definition}[equation]{Definition}

\title{The Many Classical Faces of Quantum Structures}
\author{Chris Heunen}

\begin{document}

\begin{abstract}
  Interpretational problems with quantum mechanics can be phrased precisely by only talking about empirically accessible information. This prompts a mathematical reformulation of quantum mechanics in terms of classical mechanics. 
  We survey this programme in terms of algebraic quantum theory.
\end{abstract}

\maketitle

\section{Introduction}

The mathematical formalism of quantum mechanics is open to interpretation. For example, the~possibility of deterministic hidden variables, the uncertainty principle, the measurement problem, and the reality of the wave function, are all up for debate. (The first and the last of course have rigorous restrictions: hidden variables by the Bell inequalities~\cite{bell:epr} and the Kochen--Specker theorem~\cite{kochenspecker:hiddenvariables}, discussed below, and reality of the wave function by the Pusey--Barrett--Rudolph theorem~\cite{puseybarrettrudolph:reality}.) Classical mechanics shares none of those interpretational questions. 
This article surveys a mathematical reformulation of quantum mechanics in terms of classical mechanics, intended to bring the interpretational issues with the former to a head. 
This programme proposes to replace the usual notion of state space of a quantum-mechanical system by a new one, in a way that avoids the interpretational questions above and leaves classical systems unaffected:
\begin{itemize}
  \item known obstructions to hidden variable interpretations merely say that states cannot be located with exact precision in the state space, and are circumvented via open regions of states;
  \item the uncertainty principle cannot be expressed and therefore poses no interpretational problem;
  \item the measurement problem is obviated because the new notion of state space incorporates all classical data resulting from possible measurements.
\end{itemize}
\noindent
If we also take dynamics into account, the new notion of configuration space, called an active~lattice:
\begin{itemize}
  \item yields the same predictions as traditional quantum mechanics.  
\end{itemize}
\noindent
This programme branches into a number of related themes, spread over the literature; see the extensive bibliography. The aim of this article is to bring all these active developments together to give an overview. There are hardly any new results. Instead, the novelty lies in rephrasing foundations to give an accessible, coherent, and complete overview of the current state-of-the-art. To do so, we will have to be rather brief and refer to references for many technical details. 
Nevertheless, there is a novel contribution regarding topological structure of the new notion of configuration space.
We will use an $n$-level physical system as a running example to illustrate new notions (though many results have exceptions for $n \leq 2$, and most interesting features occur in infinite dimension).
The rest of this introduction summarizes the framework and discusses four salient features, before giving an overview of the rest of this article.

\subsection{Algebraic Quantum Theory} 
The traditional formalism of quantum theory holds that the \emph{(pure) state space} is a Hilbert space $H$, that \emph{(sharp) observables} correspond to self-adjoint operators on that Hilbert space, and that \emph{(undisturbed) evolution} corresponds to unitary operators. Algebraic quantum theory instead takes the observables as primitive, and the state space is a derived notion. Self-adjoint operators combine with unitaries to give all bounded operators, and these form a so-called C*-algebra $B(H)$. However, superselection rules mandate that not all self-adjoint operators correspond to valid observables. Thus, one considers arbitrary C*-algebras, rather than only those of the form $B(H)$. Nevertheless, it turns out that any C*-algebra $A$ embeds into $B(H)$ for some Hilbert space $H$, and in that sense C*-algebra theory faithfully captures quantum theory. Finally, one could impose extra conditions on a C*-algebra, leading to so-called AW*-algebras, and W*-algebras, also known as von Neumann algebras. A good example to keep in mind is the algebra $\M_n(\C)$ of $n$-by-$n$ complex matrices, that models (the observables of) an $n$-level system, or direct sums $\M_{n_1}(\C) \oplus \cdots \oplus \M_{n_k}(\C)$.

To pass from pure to mixed states (density matrices), from sharp to unsharp observables (positive operator valued measurements), and from undisturbed evolution to including measurement (quantum channels), the traditional formalism prescribes completely positive maps. These find their natural home in the algebraic formulation. 
States of a C*-algebra $A$ can then be recovered as unital (completely) positive maps $A \to \C$. Observables with $n$ outcomes are unital (completely) positive maps $\C^n \to A$; sharp observables correspond to homomorphisms.
Evolution is described by a completely positive map $A \to A$; undisturbed evolution corresponds to a homomorphism.
Indeed, if $A=\M_n(\C)$, then states $A \to \C$ are precisely density matrices; observables $\C^n \to A$ are precisely positive operator valued measurements with $n$ outcomes; completely positive maps $A \to A$ are precisely those that map density matrices to density matrices; and homomorphisms $A \to A$ are precisely the linear functions that map pure states to pure states.

For more information on algebraic quantum theory, see~\cite{buschgrabowskilahti:operational,keyl:quantuminformation,kadisonringrose:operatoralgebras,berberian:baer,emch:foundations,davies:open,earman:superselection,redei:quantumlogic,haag:local,strocchi:quantum}.

\subsection{Gelfand Duality}
The advantage of algebraic quantum theory is that it places quantum mechanics on the same footing as classical mechanics. The \emph{(pure) state space} in classical mechanics can be any locally compact Hausdorff topological space $X$, \emph{(sharp) observables} are continuous functions $X \to \mathbb{R}$, and \emph{evolution} is given by homeomorphisms $X \to X$. This leads to the C*-algebra $C_0(X)$ of continuous complex-valued functions on $X$ vanishing at infinity; for compact $X$, we write $C(X)$. A simple example is the algebra $\C^n$, where $X$ is a discrete space with $n$ points.
Indeed, in that case there are $n$ (pure) states; (sharp) observables are precisely vectors in $\mathbb{R}^n$; and (deterministic) evolutions are just functions $n \to n$.

Again, we can pass from classical mechanics to the probabilistic setting of statistical mechanics by considering completely positive maps. States of $C(X)$ can be recovered as unital (completely) positive maps $C(X) \to \C$ as before;  pure states $x \in X$ correspond to homomorphisms. Observables with $m$ outcomes are (completely) positive maps $\C^{m} \to C(X)$, and sharp observables correspond to homomorphisms. Stochastic evolution is described by a (completely) positive map $C(X) \to C(X)$; deterministic evolution corresponds to a homomorphism.
Indeed, for $X$, the discrete space with $n$ points, states $C(X) \to \C$ are precisely probability distributions on $n$ points; observables $\C^m \to C(X)$ with $m$ outcomes are precisely $m$-tuples of probability distributions on $n$ points summing to one; sharp observables $\C^m \to C(X)$ are just functions $m \to n$; and evolutions $C(X) \to C(X)$ are simply stochastic $m$-by-$n$ matrices.

Note that multiplication in $C(X)$ is \emph{commutative}, whereas $B(H)$ was \emph{noncommutative}. Gelfand duality says that any commutative C*-algebra $C$ is of the form $C(X)$ for some compact Hausdorff space $X$, called its \emph{spectrum} and written as $\Spec(C)$. That is, $C \cong C(\Spec(C))$ and $X \cong \Spec(C(X))$. Moreover, this gives a dual equivalence of categories: if $f \colon X \to Y$ is a continuous function then $C(f) \colon C(Y) \to C(X)$ is a homomorphism, and conversely, if $f \colon C \to D$ is a homomorphism, then $\Spec(f) \colon \Spec(D) \to \Spec(C)$ is a continuous function. Thus, C*-algebra theory is often regarded as \emph{noncommutative topology}.
In~the case of a discrete space $X$ with $n$ points, this simply says that up to isomorphism $\C^n$ is the only commutative {C*-algebra} of dimension $n$, and that functions $n \to n$ are the only way to describe deterministic evolutions.

For more information, we refer to~\cite{emch:algebraic,albertiuhlmann:stochastic,landsman:topics,weaver:quantization} in addition to references above.

\subsection{Bohr's Doctrine of Classical Concepts}
To summarize, both classical systems and quantum systems are first-class citizens that can interact in the algebraic framework. Classical systems are commutative algebras $C$, and quantum systems are noncommutative ones $A$. An example interaction is measurement, given by maps $C \to A$. 
For~$n$-level systems, a measurement with $m$ outcomes is a map $\C^m \to \M_n(\C)$.
Having no superfluous outcomes in $\Spec(C)$ of the measurement corresponds to the injectivity of these maps. So the information that all possible measurements can give us about a possibly noncommutative algebra $A$ is its collection $\cC(A)$ of commutative subalgebras $C$. In other words, all empirically accessible information in a quantum system is encoded in its family of classical subsystems. This observation is known as the \emph{doctrine of classical concepts} and dates back to Bohr~\cite{bohr:einstein,heunenlandsmanspitters:topos}.
For an $n$-level system $A=\M_n(\C)$, elements of $\cC(A)$ indeed correspond to all possible measurement setups: the ways of choosing an orthonormal basis of $\C^n$ and a partition of an $n$-element set with $m$ equivalence classes for outcomes.

The main aim of this paper is to survey what can be said about the quantum structure $A$ based on its many classical faces $\cC(A)$, explaining the title. 

\subsection{The Kadison--Singer Problem}
A case in point is the long-standing but recently solved Kadison--Singer problem~\cite{kadisonsinger:extensions,marcusspielmansrivastava:kadisonsinger}. In a noncommutative C*-algebra, not all observables are compatible, in the sense that they can be measured simultaneously (without uncertainty). What can at most be measured in an experiment are those observables in a single commutative subalgebra. The best an experimenter can do is repeat the experiment to determine the values of those observables, giving a pure state of that commutative subalgebra. Ideally, this tomography procedure should determine the state of the entire system.
Indeed, there are various protocols for performing such tomography on $n$-level systems that have been experimentally verified~\cite{altepeterjameskwiat:tomography}.

The Kadison--Singer result says that this procedure indeed works in the discrete case. Let $H$ be a Hilbert space of countable dimension. Then $B(H)$ has a discrete maximal commutative subalgebra $\ell^\infty(\mathbb{N})$ consisting of operators that are diagonal in a fixed basis. The precise result is that a pure state of $\ell^\infty(\mathbb{N})$ extends \emph{uniquely} to a pure state of $B(H)$. Thus, (the state of) a quantum system is characterized by what we can learn about it from experiments, giving a positive outlook on Bohr's doctrine of classical concepts. 

\subsection{The Kochen--Specker Theorem}
Nevertheless, Bohr's doctrine of classical concepts should be interpreted carefully. It does not say that collections of states of each classical subsystem assemble to a state of the quantum system. That is ruled out by the Kochen--Specker theorem. In physical terms, local deterministic hidden variables are impossible; one cannot assign definite values to all observables of a quantum system in a noncontextual way, i.e., giving coherent states on classical subsystems. In mathematical terms, Gelfand duality does not extend to noncommutative algebras via $\cC(A)$; this will be discussed in more detail in Section~\ref{sec:kochenspecker}.
More precisely, the zero map is the only function $\M_n(\C) \to C(X)$ that restricts to homomorphisms $C \to C(X)$ for each $C \in \cC(\M_n(\C))$ when $n \geq 3$. 
That is, there is no way to assign measurement outcomes in $\R^m$ to all possible positive operator valued measures on an $n$-level system with $m$ outcomes in a consistent way.
This extends to more general noncommutative $A$ that do not contain a subalgebra $\M_2(\C)$. See~\cite{kochenspecker:hiddenvariables,redei:quantumlogic,butterfieldisham:kochenspecker}.
  
\subsection{Overview of This Article}
Section~\ref{sec:kochenspecker} continues in more depth the discussion of the structure of quantum systems from the perspective just sketched. In particular, it covers exactly how much of $A$ can be reconstructed from $\cC(A)$, and makes precise the link between the Kochen--Specker theorem and noncommutative Gelfand duality.
Section~\ref{sec:topos} shows how to interpret a quantum system $A$ as a classical system via $\cC(A)$ by changing the rules of the ambient set theory, and discusses the surrounding interesting interpretational issues.
Section~\ref{sec:domains} considers fine-graining. Increasing chains of classical subsystems give more and more information about the quantum system. We discuss $\cC(A)$ from this information-theoretic point of view, called \emph{domain theory}.
Section~\ref{sec:active} explains how to incorporate dynamics into $\cC(A)$, turning it into a so-called \emph{active lattice}. It turns out that this extra information does make $\cC(A)$ into a full invariant, from which one can reconstruct $A$. This raises interesting interpretational questions: its active lattice can be regarded as a \emph{configuration space} that completely determines a quantum system. By encoding more than static hidden variables, it circumvents the obstructions of Section~\ref{sec:kochenspecker}.
To obtain an equivalence for quantum systems like Gelfand duality did for classical ones, it thus suffices to characterize the active lattices arising this way. This is examined in Section~\ref{sec:characterization}.
Finally, Section~\ref{sec:cqm} considers to what extent the successes of the doctrine of classical concepts in the previous sections are due to the use of algebraic quantum theory, and to what extent they generalize to other formulations.


\section{Invariants}\label{sec:kochenspecker}

Bohr's doctrine of classical concepts teaches that a quantum system can only be empirically understood through its classical subsystems. These classical subsystems should therefore contain all the physically relevant information about the quantum system. 

\begin{Definition}\label{def:C}
  For a unital C*-algebra $A$, write $\cC(A)$ for its family of commutative unital C*-subalgebras $C$ (with the same unit as $A$). 
  We may think of it either as partially ordered set by inclusion, or as a diagram that remembers that the points of the partially ordered set are C*-algebras $C$.
\end{Definition}

For example, the partially ordered set $\cC(A)$ of a $2$-level system $A=\M_2(\C)$ has Hasse diagram
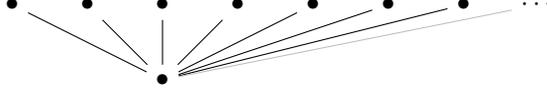
\[\begin{tikzpicture}
  \node (b) at (0,0) {$\bullet$};
  \node (1) at (-2,1) {$\bullet$}; \draw (1) to (b);
  \node (2) at (-1,1) {$\bullet$}; \draw (2) to (b);
  \node (3) at (0,1) {$\bullet$}; \draw (3) to (b);
  \node (4) at (1,1) {$\bullet$}; \draw (4) to (b);
  \node (5) at (2,1) {$\bullet$}; \draw (5) to (b);
  \node (6) at (3,1) {$\bullet$}; \draw (6) to (b);
  \node (7) at (4,1) {$\bullet$}; \draw (7) to (b);
  \node (8) at (5,1) {$\cdots$}; \draw[draw=gray!50] (8) to (b);
\end{tikzpicture}\]
with a point on the upper level for each unitary in $U(2)$.

The question is then: how does the mathematical formalism of the quantum theory of $A$ translate into terms of $\cC(A)$? 
For example, it turns out that the entropy of a state of $A$ can be reconstructed from the entropies of its restriction to $\cC(A)$~\cite{constantindoering:entropy}, see also~\cite{hamhalterturilova:orthogonalmeasures}.
Ideally, we would like to completely reconstruct $A$ from $\cC(A)$. A priori, $\cC(A)$ is merely an invariant of $A$. This section investigates how strong an invariant it is. The first step is to realize that, from $\cC(A)$, we can reconstruct $A$ as a set, as~well as operations between commuting elements. This can be made precise by the notion of a \emph{piecewise {C*-algebra}}, which is basically a C*-algebra that forgot how to add or multiply noncommuting operators.

\begin{Definition}\label{def:piecewise}
  A piecewise C*-algebra consists of a set $A$ with
  \begin{itemize}
    \item a reflexive and symmetric binary
      (\emph{commeasurability}) relation $\commeas \subseteq A \times A$;
    \item elements $0,1 \in A$;
    \item a (total) involution $* \colon A \to A$;
    \item a (total) function $\cdot \colon \mathbb{C} \times A \to A$;
    \item a (total) function $\|-\| \colon A \to \mathbb{R}$;
    \item (partial) binary operations $+, \cdot \colon  \commeas \to A$;
  \end{itemize}
  such that every set $S \subseteq A$ of pairwise commeasurable elements is contained in a set $T \subseteq A$ of pairwise commeasurable elements that forms a commutative C*-algebra under the above operations.
\end{Definition}

Of course, any commutative C*-algebra is a piecewise C*-algebra. More generally, the normal elements (those commuting with their own adjoint) of any C*-algebra $A$ form a piecewise C*-algebra. 
For an $n$-level system $A=\M_n(\C)$, the piecewise C*-algebra consists of all normal $n$-by-$n$ matrices, together with their norms and adjoints, as well as the knowledge of how commuting elements add and multiply.
Notice that $\cC(A)$ makes perfect sense for any piecewise C*-algebra $A$. To make precise how we can reconstruct the piecewise structure of $A$ from $\cC(A)$, we will use the language of \emph{category~theory}~\cite{maclane:categories}. C*-algebras, with $*$-homomorphisms between them, form a category. We can also make piecewise C*-algebras into a category with the following arrows: (total) functions $f \colon A \to B$ that preserve commeasurability and the algebraic operations, whenever defined.

The precise notion we need is that of a \emph{colimit}. Suffice to say here, a colimit, when it exists, is a universal solution that compatibly pastes together a given diagram into a single object. Thinking of $A$ as the whole and $\cC(A)$~as its parts, we would like to know whether the whole is determined by the parts. The following theorem says that $\cC(A)$  indeed contains enough information to reconstruct $A$~as a piecewise C*-algebra.

\begin{Theorem}[\cite{vdbergheunen:colim}]
  Every piecewise C*-algebra is the colimit of its commutative C*-subalgebras in the category of piecewise C*-algebras.
\end{Theorem}

This means that the diagram $\cC(A)$ determines the piecewise C*-algebra $A$: if $\cC(A)$ and $\cC(B)$ are isomorphic diagrams, then $A$ and $B$ are isomorphic piecewise C*-algebras. Moreover, the previous theorem gives a concrete way to reconstruct $A$ from $\cC(A)$. 
For the $n$-level system $A=\M_n(\C)$, this~means we can reconstruct from $\cC(A)$ the normal $n$-by-$n$ matrices, as well as sums and products of commuting ones.
An important point to note here is that the reconstruction is happening in the setting of piecewise C*-algebras. We could not have taken the colimit in the category of commutative C*-algebras instead. Indeed, one way to reformulate the Kochen--Specker theorem in terms of colimits is the following. The following reformulation might not look much like the original, but it is nevertheless equivalent, and more suited to our purposes; see also~(\cite{kochenspecker:hiddenvariables}, page 66).

\begin{Theorem}[\cite{kochenspecker:hiddenvariables,reyes:obstructing}]\label{thm:kochenspecker}
  If $n \geq 3$, then the colimit of $\cC(\M_n(\C))$ in the category of commutative C*-algebras is the degenerate, 0-dimensional, C*-algebra.
\end{Theorem}

In fact, the colimit of $\cC(A)$ degenerates for many more C*-algebras $A$ than just $\M_n(\C)$, such as any C*-algebra of the form $\M_n(B)$ for some C*-algebra $B$, or any W*-algebra that has no direct summand $\C$ or $\M_2(\C)$~\cite{bergheunen:extending,doering:kochenspecker}.

As mentioned in the introduction, Gelfand duality is a \emph{functor} from the category of commutative C*-algebras to the category of compact Hausdorff topological spaces. That is, a systematic way to assign a space to a C*-algebra, that respects functions. Interpreted physically: any classical system is determined by a configuration space in a way that respects operations on the system. The~previous theorem can be used to show that there is no such configuration space determining quantum systems---at least, if the notion of configuration space is to be a \emph{conservative extension} of the classical notion. 
The latter can be made precise as a continuous functor from the category of compact Hausdorff spaces to some category with a degenerate space like the empty set, more precisely, a strict initial object $0$.

\begin{Theorem}[\cite{bergheunen:extending}]\label{thm:nogo}
  Suppose there exist a category conservatively extending that of compact Hausdorff spaces and a functor $F$ completing the following square.
  
  \[\begin{tikzpicture}[xscale=6]
    \node (tl) at (0,1) {commutative C*-algebras};
    \node (bl) at (0,0) {C*-algebras};
    \node (tr) at (1,1) {compact Hausdorff spaces};
    \node (br) at (1,0) {?};
    \draw[->] (tl) to node[above]{$\Spec$} (tr);
    \draw[->] (bl) to node[below]{$F$} (br);
    \draw[draw=none] (tl) to node[sloped]{$\subseteq$} (bl);
    \draw[draw=none] (tr) to node[sloped]{$\subseteq$} (br);
  \end{tikzpicture}\]
  
  Then $F(\M_n(\C))=0$ for $n \geq 3$. In particular, $F$ cannot be a dual equivalence.
\end{Theorem}

Asking the functor on the right to be continuous is appropriate to model the classical limit of quantum systems converging to a classical one, because then the state space of the product of two~limiting classical systems should be computed as the classical limit of the joint quantum systems. In~fact, the proof in~\cite{bergheunen:extending} holds if the category on the bottom right has limits, and the functor on the right reflects them. However, one might still wonder if it is reasonable to ask the diagram to commute on the nose. Instead, we could ask it to commute up to a natural isomorphism. This is precisely the way out we will explore in Sections~\ref{sec:topos} and~\ref{sec:active}.

This rules out many possible quantum configuration spaces that have been proposed for the bottom right role in the square; in particular many generalized notions of topological spaces, such as sets, topological spaces themselves, pointfree topological spaces, ringed spaces, quantales, toposes, categories of sheaves, and many more~\cite{reyes:obstructing,bergheunen:extending,reyes:sheaves}. In particular, the \emph{state space} of a C*-algebra, as~discussed in the introduction, will not do for us, even though it is one of the most important tools associated with a C*-algebra~\cite{alfsenshultz:statespaces}. That explains why we deliberately talk about ``configuration spaces''. In the classical case, the two notions coincide. The previous theorem shows that serious notions of quantum configuration space must be less conservative. This points the way towards good candidates: Sections~\ref{sec:topos} and~\ref{sec:active} will cover two that do fit the bill.

The question of noncommutative extensions of Gelfand duality is also very interesting from a purely mathematical perspective. As mentioned in the introduction, C*-algebra theory can be regarded as noncommutative topology. Adding more structure than mere topology leads to \emph{noncommutative geometry}, which is a rich field of study~\cite{connes:noncommutativegeometry}. However, it takes place entirely on the algebraic side. Finding the right notion of quantum configuration space could reintroduce geometric intuition, which is usually very powerful~\cite{akemann,gileskummer}. For example, in certain cases, extensions of $\cC(A)$ can be used to compute the \emph{K-theory} of $A$, which is a way to study homotopies of the configuration space underlying $A$, that includes many local-to-global principles~\cite{desilva:ktheory}. Similarly, closed \emph{ideals} of a W*-algebra $A$, that are important because they correspond to open subsets in the classical case, are in bijection with certain piecewise ideals of $\cC(A)$~\cite{desilvabarbosa:partialideals}.

So far, we have considered $\cC(A)$ as a \emph{diagram} of parts of the whole. We finish this section by considering it as a mere partially ordered set, where we forget that elements have the structure of commutative C*-algebras. That is, we only consider the shape of how the parts fit together. This information is already enough to determine the piecewise structure of $A$, but as a \emph{Jordan algebra}. (In fact, considering $\cC(A)$ as a mere partially ordered set gives precisely the same information as considering it as a diagram~\cite{heunenlindenhovius:domains}. This justifies Definition~\ref{def:C}.) 
The self-adjoint elements of a C*-algebra form a Jordan algebra under the product $a \circ b = \tfrac{1}{2}(ab+ba)$; this even gives a so-called JB-algebra. 
In fact, any JB-algebra is a subalgebra of the direct sum of one of this form and an exceptional one, such as quaternionic matrices $\M_3(\mathbb{H})$~\cite{hancheolsenstoermer:jordan}. 
For example, the $n$-level system gives the JB-algebra of hermitian $n$-by-$n$ matrices multiplied via anticommutators.
Piecewise Jordan algebras and their homomorphisms are defined analogously to Definition~\ref{def:piecewise}. The structure of quantum observables leads naturally to the axioms of Jordan algebras~\cite{emch:foundations} (Modern mathematical physics tends to prefer C*-algebras, as their theory is slightly less complicated, and the connections to Jordan algebras are so tight anyway~\cite{hancheolsenstoermer:jordan}.) The following theorem justifies that point of view.

\begin{Theorem}[\cite{hamhalter:ordered}]
  Let $A$ and $B$ be C*-algebras. If $\cC(A)$ and $\cC(B)$ are isomorphic partially ordered sets, then $A$ and $B$ are isomorphic as piecewise Jordan algebras.
\end{Theorem}

A little more can be said. Any isomorphism $f \colon \cC(A) \to \cC(B)$ is implemented by an isomorphism $g \colon A \to B$ of piecewise Jordan algebras, in the sense that $f(C) = \{ g(c) \mid c \in C \}$. In fact, this $g$ is unique, unless $A$ is either $\C^2$ or $\M_2(\C)$. For \emph{AW*-algebras} more is true because of Gleason's theorem, that we will meet in Section~\ref{sec:active}, we can actually reconstruct the full linear structure rather than just the piecewise linear structure.
(An \emph{AW*-algebra} is a C*-algebra $A$ that has enough projections, in the sense that every $C \in \cC(A)$ is the closed linear span of its projections, and those projections work together well, in the sense that orthogonal families in the partially ordered set of projections have least upper bounds~\cite{kaplansky:awstar,berberian:baer}. See also Section~\ref{sec:active}. They are more general than W*-algebras, and much of the theory of W*-algebra generalizes to AW*-algebras, such as the type decomposition. An $n$-level system $A=\M_n(\C)$ forms a W*-algebra, and hence also an AW*-algebra.)
Type $\mathrm{I}_2$ AW*-algebras are those of the form $\M_2(C)$ for a commutative AW*-algebra $C$. AW*-algebras with a type $\mathrm{I}_2$ direct summand correspond to the exceptional case $n=2$ in the Kochen--Specker Theorem~\ref{thm:kochenspecker}. We will call them \emph{atypical}, and algebras without a type $\mathrm{I}_2$ direct summand \emph{typical}, as we will meet this exception often.
An $n$-level system is typical when $n \geq 3$.

\begin{Corollary}[\cite{doeringharding:jordan,hamhalter:dye}]\label{cor:jordan}
  Let $A$ and $B$ be typical AW*-algebras. If $\cC(A)$ and $\cC(B)$ are isomorphic partially ordered sets, then $A$ and $B$ are isomorphic as Jordan algebras.
\end{Corollary}

Whereas the C*-algebra product is associative but need not be commutative, the Jordan product is commutative but need not be associative; commutative C*-subalgebras correspond to associative Jordan subalgebras. Indeed, the previous theorem generalizes to Jordan algebras in those terms~\cite{hamhalterturilova:jordan}.

\section{Toposes}\label{sec:topos}

In this section, we consider $\cC(A)$ as a diagram. That is, we regard it as an operation that assigns to each classical subsystem $C \in \cC(A)$ of the quantum system $A$ a classical system $C$. What kind of operation is this diagram $C \mapsto C$? We can think of it as a set $S(C)$ that varies with the context $C \in \cC(A)$. 
Moreover, this \emph{contextual set} respects coarse-graining: if $C \subseteq D$, then $S(C) \subseteq S(D)$. That is, when the measurement context $C$ grows to include more observables, the information contained in the set $S(C)$ assigned to it grows along accordingly. 
For example, for an $2$-level system $A=\M_2(\C)$, this comes down to a choice of set $S(u)$ for each unitary $u \in U(2)$, that all include a fixed set $S(0)$.
Hence, these contextual sets are functors $S$ from $\cC(A)$, now regarded as a partially ordered set, to the category of sets and functions. 
The totality of all such functors forms a category. 
In fact, contextual sets form a particularly nice category, namely a \emph{topos}. 

A topos is a category that shares a lot of the properties of the category of sets and functions. In~particular, one can \emph{do mathematics inside} a topos: we may think about objects of a topos as sets, that we may specify and manipulate using logical formulae. Of course, this internal perspective comes with some caveats. Most notably, if a proof is to hold in the internal language of any topos, it has to be \emph{constructive}: we are not allowed to use the axiom of choice or proofs by contradiction, and have to be careful about real numbers. We cannot go into more detail here, but for more information on topos theory, see~\cite{johnstone:elephant}.

One particular object of interest in the topos of contextual sets over $\cC(A)$ is our canonical contextual set $C \mapsto C$. It turns out that, according to the logic of the topos of contextual sets, this object is a \emph{commutative} C*-algebra.

\begin{Theorem}[\cite{heunenlandsmanspitters:topos}]\label{thm:bohrification}
  Let $A$ be a C*-algebra. In the topos of contextual sets over $\cC(A)$, the canonical contextual set $C \mapsto C$ is a commutative C*-algebra.
\end{Theorem}

This procedure is called \emph{Bohrification}:
\begin{enumerate}
  \item Start with a quantum system $A$.
  \item Change the logical rules of set theory by moving to the topos of contextual sets over $\cC(A)$.
  \item The quantum system $A$ turns into a classical one given by the canonical contextual set $C \mapsto C$.
\end{enumerate}
See also~\cite{landsman:bohrification}.

Thus, we may study the quantum system $A$ as if it were a classical system. 
Of course, we lose the same information as in the previous section. For example, we can only hope to reconstruct the Jordan structure of $A$ from the contextual set $C \mapsto C$. Nevertheless, placing it in a topos of its peers opens up many possibilities. In particular, we may try to find a configuration space inside the topos. It turns out that Gelfand duality can be formulated so that its proof is constructive, and hence applies inside the topos. This involves talking about \emph{locales} rather than topological spaces. We may think of a locale as a topological space that forgot it had points. 

More precisely, a locale may be thought of as the partially ordered family of open sets of a topological space, but without a carrier set of points.
Most of topology can be formulated to work for locales as well. Again, we cannot go into more detail here, but for more information on locales see~\cite{johnstone:stonespaces}.

\begin{Corollary}[\cite{banaschewskimulvey06}]
  Let $A$ be a C*-algebra. In the topos of contextual sets over $\cC(A)$, there is a compact Hausdorff locale $X$ such that the canonical contextual set is of the form $C(X)$.
\end{Corollary}

For example, if $A$ is the $2$-level system $\M_2(\C)$, then $X$ is the contextual set $S$ that assigns to $u \in U(2)$ the orthonormal basis of $\C^2$ corresponding to $u$, and that assigns to $0$ the zero vector in $\C^2$, where $S(u)$ locally carries the structure of a 2-element discrete space, and $S(0)$ carries the structure of a 1-element discrete space.
We will call this locale $X$ the \emph{spectral contextual set}. In general, it is not just the contextual set $C \mapsto \Spec(C)$. However, it does resemble that if we think about \emph{bundles} instead of contextual sets~\cite{spittersvickerswolters:geometric,fauserraynaudvickers:bundles}:
a bundle is a map of locales into the locale of ideals of $\cC(A)$, and by restricting the intuitionistic logic of a topos further to so-called geometric logic, the bundle corresponding to the spectral contextual set does have fibre $\Spec(C)$ over $C$.
Also, if we reverse the partial order on $\cC(A)$, the assignment $C \mapsto \Spec(C)$ plays the role of the canonical contextual set. So there are two approaches:
\begin{itemize}
	\item Either one uses $\cC(A)$; the canonical contextual set $C \mapsto C$ is a commutative C*-algebra, and the spectral contextual set $X$ does not take a canonical form~\cite{heunenlandsmanspitters:topos,heunenlandsmanspitters:bohrification,caspersheunenlandsmanspitters:intuitionistic,heunenlandsmanspitters:logic,wolters:topos,nuiten:aqft}.
	\item Or one uses the opposite order; the spectral contextual set $X$ is a locale of the canonical form $C \mapsto \Spec(C)$, and the commutative C*-algebra $C(X)$ does not take a canonical form~\cite{doeringisham:topos,doeringisham:thing,doeringisham:neorealist,flori:topos}.
\end{itemize}

For a comparison, see~\cite{wolters:toposes}. For this overview article, the choice of direction does not matter so much. In any case, $X$ is an object inside the topos of contextual sets, and as such we may reason about it as a locale. In particular, we may wonder whether it is a topological space, that is, whether it does in fact have enough points. It turns out that the Kochen--Specker Theorem~\ref{thm:kochenspecker} can be reformulated as saying that not only does $X$ not have enough points, in fact it has no points at all. 
In terms of bundles: the canonical bundle has no global sections.
This illustrates the need for locales rather than topological~spaces.

\begin{Proposition}[\cite{butterfieldisham:kochenspecker}]\label{prop:pointfree}
  Let $A$ be a C*-algebra satisfying the Kochen--Specker Theorem~\ref{thm:kochenspecker}. In the topos of contextual sets over $\cC(A)$, the spectral contextual set has no points.
\end{Proposition}

Thus, Bohrification turns a quantum system $A$ into a locale $X$ inside the topos of contextual sets over $\cC(A)$. There is an equivalence between locales $X$ inside such a topos over $\cC(A)$, and certain continuous functions from a locale $\Spec(A)$ to $\cC(A)$ \emph{outside} the topos~\cite{joyaltierney:galois}. This gives a way to cut out the whole topos detour, and assign to the quantum system $A$ a configuration space that we will temporarily call $\Spec(A)$ for the rest of this section.

\begin{Proposition}[\cite{heunenlandsmanspitterswolters:gelfand}]
  For any C*-algebra $A$, the internal locale $X$ is determined by a continuous function from some locale $\Spec(A)$ to $\cC(A)$.
\end{Proposition}

In many cases, $\Spec(A)$ will in fact have enough points, i.e., will be a topological space~\cite{heunenlandsmanspitterswolters:gelfand,wolters:toposes}--- despite Proposition~\ref{prop:pointfree}. The construction $A \mapsto \Spec(A)$ circumvents the obstruction of Theorem~\ref{thm:nogo} for several reasons. First, when the C*-algebra $A$ is commutative, $\Spec(A)$ turns out to be a locale based on $\cC(A)$, rather than on $A$ itself; therefore what we are currently denoting by $\Spec(A)$ does not match the Gelfand spectrum of $A$. Second, the construction $A \mapsto \Spec(A)$ is only partially functorial: if we regard $\cC(A)$ as a locale, the construction only respects functions that reflect commutativity~\cite{vdbergheunen:colim}, and to get functorality we have to regard $\cC(A)$ as a \emph{localed topos}, that is, a topos with a locale in it~\cite{vdbergheunen:colim:erratum}.

We can only touch on it briefly here, but one of the main features of building the topos of contextual sets over $\cC(A)$ and distilling the configuration space $\Spec(A)$ is that they encode a contextual \emph{logic}. This logic is intuitionistic, and therefore very different from traditional \emph{quantum logic}~\cite{caspersheunenlandsmanspitters:intuitionistic}. The latter concerns the set $\Proj(A)$ of yes--no questions on the quantum system $A$; more precisely, the set of sharp observables with two outcomes. These correspond to \emph{projections}: $p \in A$ satisfying $p^2=p=p^*$. They are partially ordered by $p \leq q$ when $pq=p$, which should be read as saying that $p$ implies $q$. Similarly, least upper bounds in $\Proj(A)$ are logical disjunctions~\cite{redei:quantumlogic}. 
In an $n$-level system $A=\M_n(\C)$, projections correspond to subspaces of $\C^n$, regarded logically as the set of (pure) states where the proposition is true; the order becomes inclusion of subspaces; and the disjunction of subspaces is their linear span. 
AW*-algebras $A$ are determined to a great extent by their projections, and indeed the quantum logic $\Proj(A)$ carries precisely the same amount of information as $\cC(A)$~\cite{heunen:cccc}.
For more information about this topos-theoretic approach to quantum logic, we refer to~\cite{heunenlandsmanspitters:bohrification,heunenlandsmanspitters:topos,heunenlandsmanspitters:logic,caspersheunenlandsmanspitters:intuitionistic,doeringisham:topos,doeringisham:thing,doeringisham:neorealist,spittersvickerswolters:geometric,wolters:topos}.

To connect contextual sets to probabilities and the Born rule, we have to translate states of $A$ into some notion based on the spectral contextual set $X$, and observables of $A$ into some notion based on the canonical contextual set $C \mapsto C$. For the latter, one has to resort to approximations, as not every $a \in A$ will be present in each $C \in \cC(A)$; this process is sometimes called \emph{daseinisation}~\cite{doeringisham:thing}. The former has a satisfying solution in terms of \emph{piecewise states}: piecewise linear (completely) positive maps $A \to \C$.

\begin{Theorem}[\cite{heunenlandsmanspitters:bohrification,spitters:measurement,butterfieldisham:kochenspecker}]\label{thm:piecewisestates}
  There is a bijective correspondence between piecewise states on an AW*-algebra $A$, and states of the canonical contextual set $C \mapsto C$ inside the topos of contextual sets over $\cC(A)$. 
\end{Theorem}

(The cited references consider W*-algebras, but the proof holds for AW*-algebras because Corollary~\ref{cor:piecewiseJordan} does so, see Section~\ref{sec:active}. The same goes for the references in Corollary~\ref{cor:piecewisestates}.) By Gleason's theorem (see Section~\ref{sec:active}), we can say more for AW*-algebras. 
See also~\cite{hamhalterturilova:orthogonalmeasures}.

\begin{Corollary}[\cite{degroote:states,doering:measures}]\label{cor:piecewisestates}
  There is a bijective correspondence between states of a typical AW*-algebra $A$, and states of the canonical contextual set $C \mapsto C$ inside the topos of contextual sets over $\cC(A)$.
\end{Corollary}

In the $n$-level system $A=\M_n(\C)$ for $n\geq 3$, this means that $n$-by-$n$ density matrices correspond precisely to a choice of probability distribution over $m$ points that is consistent over all unitaries $u \in U(n)$ and partitions of $n$ points into $m$ equivalence classes.

Combining daseinisation with the above results gives rise to a contextual Born rule, justifying the Bohrification procedure of Theorem~\ref{thm:bohrification}~\cite{fauserraynaudvickers:bundles}. Summarizing, we can formulate the physics of the quantum system $A$ completely in terms of $\cC(A)$ and its topos of contextual sets, and work within there as if dealing with a classical system. 

To end this section, let us mention some other related work. The ``amount of nonclassicality'' of the contextual logic discussed of $A$ measures the computational power of the quantum system $A$~\cite{loveridgedridiraussendorf:topos}. For~philosophical aspects of Bohrification and related constructions, see~\cite{heunenlandsmanspitters:tovariance,eppersonzafiris:realism}. Similar contextual ideas have been used to model quantum numbers~\cite{adelmancorbett:sheaf}. 
Transfering C*-algebras between different toposes has been used successfully before in so-called Boolean-valued analysis~\cite{takeuti:cstarboolean,ozawa:transfer,ozawa:typeiawstar}.
Finally, contextuality and the Kochen--Specker theorem can be formulated more generally than in algebraic quantum theory~\cite{abramskybrandenburger:sheaf}.

\section{Domains}\label{sec:domains}

The partially ordered set $\cC(A)$ of empirically accessible classical contexts $C$ of a quantum system $A$ embodies \emph{coarse-graining}. As in the introduction, we think of each $C \in \cC(A)$ as consisting of compatible observables that we can measure together in a single experiment. Larger experiments, involving more observables, should give us more information, and this is reflected in the partial order: if $C \subseteq D$, then $D$ contains more observables, and hence provides more information. If $A$ itself is noncommutative, the best we can do is approximate it with larger and larger commutative subalgebras $C$. This sort of informational approximation is studied in computer science under the name \emph{domain theory}~\cite{abramskyjung:domaintheory,gierzetal:domains}. This~section discusses the domain-theoretic properties of $\cC(A)$. Domain theory is mostly concerned with partial orders where every element can be approximated by finite ones, as those are the ones we can measure in practice, leading to the following definitions. 

\begin{Definition}
 A partially ordered set $(\cC,\leq)$ is directed complete when every ascending chain $\{D_i\}$ has a least upper bound $\bigvee_i D_i$. An element $C$ approximates $D$, written $C \ll D$, when $D \leq \bigvee_i D_i$ implies $C \leq D_i$ for any chain $\{D_i\}$ and some $i$. An element $C$ is finite when $C \ll C$.
A continuous domain is a directed complete partially ordered set, every element of which satisfies $D=\bigvee \{C \mid C \ll D\}$. An algebraic domain is a directed complete partially ordered set, every element of which is approximated by finite ones: $D=\bigvee \{ C \mid C \ll C \leq D \}$.
\end{Definition}

\begin{Lemma}[\cite{doeringbarbosa:domains,spitters:measurement}]
  If $A$ is a C*-algebra, then $\cC(A)$ is a directed complete partially ordered set, in which $\bigvee_i C_i$ is the norm-closure of $\bigcup_i C_i$.
\end{Lemma}

We saw in Section~\ref{sec:kochenspecker} that $\cC(A)$ captures precisely the structure of $A$ as a (piecewise) Jordan algebra. Order-theoretic techniques give an alternative proof of Corollary~\ref{cor:jordan}. First, we can recognize the dimension of $A$ from $\cC(A)$. Recall that a partially ordered set is \emph{Artinian} when: every nonempty subset has a minimal element; every nonempty filtered subset has a least element; every descending sequence $C_1 \geq C_2 \geq \cdots$ eventually becomes constant. The dual notion, satisfying an ascending chain condition, is called \emph{Noetherian}.

\begin{Proposition}[\cite{lindenhovius:artinian}]
  A C*-algebra $A$ is finite-dimensional if and only if $\cC(A)$ is Artinian, if and only if $\cC(A)$ is~Noetherian.
\end{Proposition}
Indeed, in an $n$-level system $A=\M_n(\C)$, elements $C \in \cC(A)$ correspond to a choice of unitary $u \in U(n)$ and a partition of $n$ points into $m$ equivalence classes. Because $C \subseteq D$ when the partition for $D$ is finer than that for $C$, the partially ordered set $\cC(A)$ can only have strictly increasing chains of length at most $n$.

By the Artin--Wedderburn theorem, we know that any finite-dimensional C*-algebra $A$ is a finite direct sum of matrix algebras $\M_{n_i}(\C)$. It is therefore specified up to isomorphism by the numbers $\{n_i\}$, which we can extract from the partially ordered set $\cC(A)$. A partially ordered set $\cC$ is called directly indecomposable when $\cC=\cC_1 \times \cC_2$ implies that either $\cC_1$ or $\cC_2$ is a singleton set.

\begin{Proposition}[\cite{lindenhovius:artinian,lindenhovius:thesis}]
  If $A=\bigoplus_{i=1}^n \M_{n_i}(\C)$, then the C*-subalgebras $\M_{n_i}(\C)$ correspond to directly indecomposable partially ordered subsets $\cC_i$ of $\cC(A)$, and furthermore $n_i$ is the length of a maximal chain in $\cC_i$.
\end{Proposition}

The previous proposition does not generalize to arbitrary C*-algebras, which need not have a decomposition as a direct sum of factors. One might expect that $\cC(A)$ is a domain when $A$ is \emph{approximately finite-dimensional}, as this would match with the intuition of approximation using practically obtainable information. However, there also needs to be a large enough supply of projections for this to work; see also Section~\ref{sec:topos}. It turns out that the correct notion is that of \emph{scattered} C*-algebras~\cite{jensen:scattered}, that is, C*-algebras $A$ for which every positive map $A \to \C$ is a sum of pure ones.
The $n$-level system $A=\M_n(C)$ is scattered.

\begin{Theorem}[\cite{heunenlindenhovius:domains}]
  A C*-algebra $A$ is scattered if and only if $\cC(A)$ is a continuous domain if and only if $\cC(A)$ is an algebraic domain.
\end{Theorem}

Compare this to the situation using commutative W*-subalgebras $\cV(A)$ of a W*-algebra $A$: $\cV(A)$ is a continuous or algebraic domain only when $A$ is finite-dimensional~\cite{doeringbarbosa:domains}.
Connecting back to Theorem~\ref{thm:piecewisestates} and Corollary~\ref{cor:piecewisestates}, let us notice that $\C$ can also be regarded as a domain using the interval topology: smaller intervals approximate an ideal complex number better than larger ones. Moreover, (piecewise) states $A \to \C$ respect such approximations: the induced functions from $\cC(A)$ to the interval domain on $\C$ are \emph{Scott continuous}~\cite{doeringbarbosa:domains,spitters:measurement}.

There are several topologies with which one could adorn $\cC(A)$. As any partially ordered set, it carries the order topology. We have just mentioned the Scott topology on directed complete partially ordered sets. For the purposes of information approximation that we are interested in, there is the \emph{Lawson topology}, which refines both the Scott topology and the order topology. If the domain is continuous, the topological space will be Hausdorff. The topological space will be compact for so-called FS-domains, which $\cC(A)$ happens to be.

\begin{Corollary}[\cite{gierzetal:domains}]\label{cor:lawson}
  For a scattered C*-algebra $A$, the Lawson topology makes $X=\cC(A)$ compact Hausdorff.
  Hence to each scattered C*-algebra $A$ we may assign a commutative C*-algebra $C(X)$.
\end{Corollary}

The assignment $A \mapsto C(\cC(A))$ is not functorial, does not leave commutative C*-algebras invariant, and of course only works for scattered C*-algebras $A$ in the first place~\cite{heunenlindenhovius:domains}. Hence there is no contradiction with Theorem~\ref{thm:nogo}.

One can also furnish $\cC(A)$ with a topology inspired by the topology of $A$ itself.
We will use the topology induced by the following variation on the \emph{Hausdorff metric}; similar variations are named after Banach-Mazur, Kadets~\cite{kaltonostrovskii:distances}, Gromov-Hausdorff, Effros-Mar{\'e}chal~\cite{haagerupwinslow:effrosmarechal}, and Kadison--Kastler~\cite{kadisonkastler:perturbations}
. See~also~\cite{chetcutihamhalterweber:ordertopology}.
Define the distance between $C,D \in \cC(A)$ to be 
\[
 d(C,D) = 
 \max \big(
  \adjustlimits\sup_{\substack{c \in C \\ \|c\|\leq 1}} \inf_{\substack{d \in D \\ \|d\|\leq 1}} \|c-d\| ,\;\;
  \adjustlimits\sup_{\substack{d \in D \\ \|d\|\leq 1}} \inf_{\substack{c \in C \\ \|c\|\leq 1}} \|c-d\| 
  \big)\text.
\]

Now if $C$ and $D$ are generated by projections $p$ and $q$, and $A$ is represented on a Hilbert space $H$,~then
\[
  \|p-q\|
  = \sup_{\substack{x \in H \\ \|x\| \leq 1}} \|p(x)-q(x)\|
  = \sup_{\substack{x \in p(H) \\ \|x\| \leq 1}} \|x-q(x)\|
  = \adjustlimits\sup_{\substack{x \in p(H) \\ \|x\| \leq 1}} \inf_{\substack{y \in q(H) \\ \|y\|\leq 1}} \|x-y\|
\]
is the Hausdorff distance between $p(H)$ and $q(H)$. It follows that the distance between $C$ and $D$ is $\max(\|p-q\|,\|(1-p)-q\|, \|p-(1-q)\|, \|(1-p)-(1-q)\|) = \max ( \|p-q\|, \|(1-p)-q\|)$.
This~topology on $\cC(A)$ matches the case of the 2-level system $A=\M_2(\C)$, where $\cC(A)$ is in bijection with the one-point compactification of the real projective plane $\mathbb{R}\mathbb{P}^2$~\cite{fauserraynaudvickers:bundles}.


\section{Dynamics}\label{sec:active}

So far, we have only considered kinematics of the quantum system $A$, by looking for configuration spaces based on $\cC(A)$. It is clear, however, that $\cC(A)$ in itself is not enough to reconstruct all of $A$. For a counterexample, observe that any C*-algebra $A$ has an opposite C*-algebra $A\op$ in which the multiplication is reversed. Clearly, $\cC(A)$ and $\cC(A\op)$ are isomorphic as partially ordered sets, but there exist C*-algebras $A$ that are not isomorphic to $A\op$ as C*-algebras~\cite{connes:factornotantiisomorphic}. So we need to add more information to $\cC(A)$ to be able to reconstruct $A$ as a C*-algebra, which is the topic of this section. To do so, we bring dynamics into the picture. For motivation of why dynamics and configuration spaces should go together, see also~\cite{spekkens:kinematics}.

We begin by viewing dynamics as a time-dependent group of evolutions. The traditional view is that the 1-parameter group consists of unitary evolutions of the Hilbert space. 
For an $n$-level system, these 1-parameter groups are continuous homomorphisms $\mathbb{R} \to U(n)$.
In algebraic quantum theory, it~becomes a 1-parameter group of isomorphisms $A \to A$ of the C*-algebra. 

  The group $\Aut(A)$ inherits the pointwise norm topology from $A$, that has subbasis
  \[
    \{g \in \Aut(A) \mid \forall a \in S \colon \|f(a)-g(a)\|<\varepsilon>\|f(a)-g(1-a)\|\}
  \]
  for $f \in \Aut(A)$, $\varepsilon>0$, and $S \subseteq A$ finite,
  and makes conjugation $U(A) \to \Aut(A)$ continuous~\cite{moffat:automorphisms}.
We~can similarly consider 1-parameter groups of isomorphisms $\cC(A) \to \cC(A)$ of partially ordered sets.

  Similarly, $\Aut(\cC(A))$ becomes a topological group with subbasis
  \[
    \{g \in \Aut(\cC(A)) \mid \forall C \in S\colon d(f(C),g(D))<\varepsilon \}
  \]
  for $f \in \Aut(\cC(A))$, $\varepsilon>0$, and finite sets $S$ of atoms of $\cC(A)$.

\begin{Definition}
Let $A$ be a C*-algebra. A 1-parameter group on $A$ is a continuous injection $\varphi \colon \R \to \Aut(\emph{A})$, that assigns to each $t \in \R$ an isomorphism $\varphi_t \colon A \to A$ of C*-algebras, satisfying $\varphi_0=1$ and $\varphi_{t+s}=\varphi_t \circ \varphi_s$.
 A 1-parameter group on $\cC(A)$ is a continuous injection $\alpha \colon \R \to \Aut(\cC(A))$, that assigns to each $t \in \R$ an isomorphism $\alpha_t \colon \cC(A) \to \cC(A)$ of partially ordered sets, satisfying $\alpha_{t+s} = \alpha_t \circ \alpha_s$.
\end{Definition}

The following theorem shows that both notions in fact coincide. A \emph{factor} is an algebra with trivial center, that is, a single superselection sector: the $n$-level system $\M_n(\C)$ is a factor, but $\M_m(\C) \oplus \M_n(\C)$ is not, because its center is two-dimensional. More precisely, the following theorem shows that the only freedom between the two notions in the previous definition lies in permutations of the center, because $\Aut(A) \simeq \Aut(\cC(A))$ for typical AW*-factors.

\begin{Theorem}[\cite{hamhalterturilova:automorphisms,doering:flows}]
  Let $A$ be a typical AW*-factor. 
  Any 1-parameter group on $\cC(A)$ is induced by a 1-parameter group on $A$, and vice versa.
\end{Theorem}

So C*-dynamics of $A$ can be completely justified in terms of $\cC(A)$.
This also justifies our choice of the topology on $\cC(A)$ induced by the Hausdorff metric. See also~\cite{heunenlindenhovius:domains:lmcs}.
Equilibrium states are described in algebraic quantum theory by \emph{Kubo--Martin--Schwinger} states, and these can be described in terms of $\cC(A)$ as well, see~\cite{gelounflori:kms}.

We now switch gear. By Stone's theorem, 1-parameter groups of unitaries $e^{ith}$ in certain W*-algebras correspond to self-adjoint (possibly unbounded) observables $h$. Thus, we may forget about the explicit dependence on a time parameter and consider single self-adjoint elements of C*-algebras. In fact, we will mostly be interested in \emph{symmetries}: self-adjoint unitary elements $s=s^*=s^{-1}$. 

Symmetries are tightly linked to projections. Every projection $p$ gives rise to a symmetry $1-2p$, and every symmetry $s$ comes from a projection $(1-s)/2$. As they are unitary, the symmetries of a C*-algebra $A$ generate a subgroup $\Sym(A)$ of the unitary group. For a commutative C*-algebra $A=C(X)$, symmetries compose, so that $\Sym(A)$ consists of symmetries only. For an $n$-level system $A=\M_n(\C)$, it turns out that $\Sym(A)$ consists of those unitaries $u \in U(n)$ whose determinant is $1$ or~$-1$. This `orientation' is what we will add to $\cC(A)$ to make it into a full invariant of $A$. See also~\cite{alfsenshultz:orientation}. 

Having enough symmetries means having enough projections. Therefore, we now consider AW*-algebras rather than general C*-algebras.  For commutative AW*-algebras $C(X)$, the Gelfand spectrum $X$ is not just compact Hausdorff, but \emph{Stonean}, or \emph{extremally disconnected}, in the sense that the closure of an open set is still open. (For comparison, the Lawson topology in Corollary~\ref{cor:lawson} is totally disconnected, in the sense that connected components are singleton sets, which is weaker than~Stonean).

Gelfand duality restricts to commutative AW*-algebras and Stonean spaces. Another way to put this is to say that the projections $\Proj(A)$ of a commutative AW*-algebra $A$ form a \emph{complete Boolean algebra}, and~vice versa, every complete Boolean algebra gives a commutative AW*-algebra. 
The~appropriate homomorphisms between AW*-algebras are \emph{normal}, meaning that they preserve least upper bounds of projections \cite{heunenreyes:activelattice}.
There are versions of Definition~\ref{def:piecewise} for piecewise AW*-algebras, and~piecewise complete Boolean algebras, too~\cite{heunenreyes:activelattice}. One could also define a piecewise Stonean space, but the following lemma suffices here.

\begin{Lemma}[\cite{heunenreyes:activelattice}\label{lem:piecewiseStone}]
  The category of piecewise complete Boolean algebras and the category of piecewise AW*-algebras are equivalent.
\end{Lemma}

The \emph{orthocomplement} $p \mapsto 1-p$ makes sense for the projections $\Proj(A)$ of any C*-algebra $A$.
We~can now make precise what equivariance under symmetries achieves: it makes the difference between being able to recover Jordan structure and C*-algebra structure.

\begin{Proposition}[\cite{heunenreyes:activelattice,hamhalter:dye}]\label{prop:equivariance}
  Let $A$ and $B$ be typical AW*-algebras, 
  and suppose that $f \colon \Proj(A) \to \Proj(B)$ preserve least upper bounds and orthocomplements.
  Then $f$ extends to a Jordan homomorphism $A \to B$. 
  It~extends to a homomorphism if additionally $f\big((1-2p)(1-2q)\big)=\big(1-2f(p)\big)\big(1-2f(q)\big)$.
\end{Proposition}

To arrive at a good configuration space for $A$, we can package all this information up. We saw that $\Proj(A)$ embedded in $\Sym(A)$. Conversely, $\Sym(A)$ acts on $\Proj(A)$: a symmetry $s$ and a projection $p$ give rise to a new projection $sps$. In this way, $\Proj(A)$ \emph{acts on itself}, and we may forget about $\Sym(A)$. Including this action leads to the notion of an \emph{active lattice} $\AProj(A)$. 
More precisely, an active lattice consists of a complete orthomodular lattice $P$, a group $G$ generated by $1-2p$ for $p \in P$ within the unitary group of the piecewise AW*-algebra $A(P)$ with projections $P$, and an action of $G$ on $P$ that becomes conjugation on $A(P)$.
The active lattice of an $n$-level system $A=\M_n(\C)$ has, for $P$, the lattice of subspaces of $\C^n$; for $G$, the group $\{u \in U(n) \mid \det(u)=\pm 1\}$; the injection $P \to G$ sends $V \subseteq \C^n$ to the reflection in $V$; and $u \in G$ acts on $V \in P$ as $uVu^*=\{uvu^* \mid v \in V \} \subseteq \C^n$. 
For morphisms of active lattices, we refer to~\cite{heunenreyes:activelattice}, but let us point out that thanks to Lemma~\ref{lem:piecewiseStone} they can be phrased in terms of projections alone, just like the above definition of the active lattice itself.
See also~\cite{chevalier:automorphisms}.
We can now make precise that we can reconstruct an AW*-algebra $A$ from its active lattice $\AProj(A)$. Up to now, we have mostly considered reconstructions of the form ``if some structures based on $A$ and $B$ are isomorphic, then so are $A$ and $B$''. The following theorem gives a much stronger form of reconstruction. Recall that a functor $F$ is \emph{fully faithful} when it gives a bijection between morphisms $A \to B$ and $F(A) \to F(B)$.

\begin{Theorem}[\cite{heunenreyes:activelattice}]\label{thm:activelattice}
  The functor that assigns to an AW*-algebra $A$ its active lattice $\AProj(A)$ is fully faithful.
\end{Theorem}

It follows immediately that if $A$ and $B$ are AW*-algebras with isomorphic active lattices $\AProj(A) \cong \AProj(B)$, then $A\cong B$ are isomorphic AW*-algebras. That is, its active lattice completely determines an AW*-algebra. We can therefore think of them as configuration spaces. As mentioned before, $\Proj(A)$ contains precisely the same information as $\cC(A)$, so we could phrase active lattices in terms of $\cC(A)$ as well. This configuration space circumvents the obstruction of Theorem~\ref{thm:nogo}, because active lattices are not a conservative extension of the ``passive lattices'' coming from compact Hausdorff spaces. Another thing to note about the previous theorem is that it has no need to except atypical cases such as $\M_2(\C)$. Finally, let us point out that functoriality of $A \mapsto \AProj(A)$ is nontrivial~\cite{heunenreyes:diagonal}.

To get a good notion of configuration space for general quantum systems, we would eventually like to pass from AW*-algebras to C*-algebras. One way to think about this step is as refining an underlying carrying set to a topological space, that is, moving from algebras $\ell^\infty(X)$ of all (bounded) functions on the set $X$ to algebras $C(X)$ of continuous functions on the topological space $X$. One might hope that AW*-algebras or W*-algebras play the former role in a noncommutative generalization, and to some extent this works~\cite{kornell:collections,kornell:vstar}. Unfortunately, the Kadison--Singer problem raises rigorous obstructions to the most obvious noncommutative generalization of such a ``discretization'' of C*-algebras to AW*-algebras~\cite{heunenreyes:discretization}.

Nevertheless, AW*-algebras are pleasant to work with. Their theory is entirely algebraic, whereas the theory of (commutative) W*-algebras involves a good deal of measure theory. For example, Gelfand spectra of commutative AW*-algebras are Stonean spaces, whereas Gelfand spectra of commutative W*-algebras are so-called hyperstonean spaces; they additionally have to satisfy a measure-theoretic condition that seems divorced from topology. A similar downside occurs with projections: the projection lattice of a commutative W*-algebra is not just a complete Boolean algebra, it additionally has to satisfy a measure-theoretic condition. In particular, projections of an enveloping AW*-algebra should correspond to certain ideals in a C*-algebra, without needing measure-theoretic intricacies.

Much of the theory of W*-algebra finds its natural home in AW*-algebras at any rate. As a case in point, consider \emph{Gleason's theorem}. It states that any probability measure on $\Proj(\M_n(\C))$ extends to a positive linear function $\M_n(\C) \to \C$ when $n>2$. Roughly speaking, any quantum probability measure $\mu$ is of the form $\mu(p) = \mathrm{Tr}(\rho p)$ for some density matrix $\rho$. 
In the algebraic formulation, any probability measure $\Proj(A) \to \C$ extends to a state $A \to \C$, for an $n$-level system $A=\M_n(\C)$~\cite{mackey}. One~can replace $A$ by an arbitrary W*-algebra, and one can even replace $\C$ by an arbitrary operator algebra $B$~\cite{buncewright:mackeygleason,hamhalter:measures}. Thanks to Proposition~\ref{prop:equivariance}, Gleason's theorem generalizes to many typical AW*-algebras $A$, such as those of so-called homogeneous type I, and those generated by two projections, which leads to the following corollary, that supports many results in Sections~\ref{sec:kochenspecker} and~\ref{sec:topos}.

\begin{Corollary}[\cite{hamhalter:dye}]\label{cor:piecewiseJordan}
  Any normal piecewise Jordan homomorphism between typical AW*-algebras is a Jordan~homomorphism.
\end{Corollary}

\section{Characterization}\label{sec:characterization}

Now that we have seen that most of the algebraic quantum theory of $A$ can be phrased in terms of $\cC(A)$ only, let us try to axiomatize $\cC(A)$ itself. Given any partially ordered set, when is it of the form $\cC(A)$ for some quantum system $A$? An answer to this question would, for example, make Theorem~\ref{thm:activelattice} into an equivalence of categories, bringing configuration spaces for quantum systems on a par with Gelfand duality for classical systems. An axiomatization would also open up the possibility of generalizations, that might go beyond algebraic quantum theory.

We start with the classical case, of commutative C*-algebras $C(X)$. By Gelfand duality, any $C \in \cC(C(X))$ corresponds to a quotient $X \slash \smash{\sim}$. In turn, the equivalence relation corresponds to a \emph{partition} of $X$ into equivalence classes. Partitions are partially ordered by refinement: if $C \subseteq D$, then any equivalence class in the partition corresponding to $D$ is contained in an equivalence class of the partition corresponding to $C$. Hence axiomatizing $\cC(C(X))$ comes down to axiomatizing \emph{partition lattices}, and this has been well-studied, both in the finite-dimensional case~\cite{birkhoff:lattices,stonesiferbogart:partitionlattices}, and in the general case~\cite{firby:compactifications1}. The list of axioms is too long to reproduce here, but let us remark that it is based on a definition of \emph{points} of the partition lattice. In the case of a finite partition lattice, the points are simply the \emph{atoms}, that is, the minimal nonzero elements. 
So for a classical system $\C^n$ with $n$ states, the elements of the partition lattice $\cC(\C^n)\op$ are the ways to partition a set of $n$ points into $m$ equivalence classes; the~atoms put two of the $n$ points in an equivalence class and all the others in their own equivalence class of one point each.
The other axioms are geometric in nature.

\begin{Lemma}[\cite{heunen:cccc}]\label{lem:partition}
  A partially ordered set is isomorphic to $\cC(C(X))$ for a compact Hausdorff space $X$ if and only if it is opposite to a partition lattice whose points are in bijection with $X$.
\end{Lemma}

Thanks to (a variation of) Lemma~\ref{lem:piecewiseStone}, the same strategy applies to piecewise Boolean algebras $B$. Write $\cC(B)$ for the partially ordered set of Boolean subalgebras of $B$. The \emph{downset} of an element $D$ of a partially ordered set consists of all elements $C \leq D$. In fact, the idea that any quantum logic (piecewise Boolean algebra) should be seen as many classical sublogics (Boolean algebras) pasted together, is not new, and drives much of the research in that area~\cite{gudder,finch,vdbergheunen:colim,hughes,scheibe:logical}.

\begin{Theorem}[\cite{heunen:piecewiseboolean}]
  A partially ordered set is isomorphic to $\cC(B)$ for a piecewise Boolean algebra $B$ if and only~if:
  \begin{itemize}
    \item it is an algebraic domain;
    \item any nonempty subset has a greatest lower bound; 
    \item a set of atoms has an upper bound whenever each pair of its elements does;
    \item the downset of each compact element is isomorphic to the opposite of a finite partition lattice.
  \end{itemize}
\end{Theorem}

In the case of a classical system with $n$ states, $B$ is the powerset of $n$ points, and the above conditions merely say that $\cC(B)\op$ is a partition lattice.

Just like in Section~\ref{sec:topos}, if we consider $\cC(B)$ as a diagram rather than a mere partially ordered set, we can reconstruct $B$. Starting from just the partially ordered set $\cC(B)$, the same issues surface as in Sections~\ref{sec:kochenspecker} and~\ref{sec:active}, about Jordan structure verses full algebra structure. In the current piecewise Boolean setting, it can be solved neatly by adding an \emph{orientation} to $\cC(B)$~\cite{heunen:piecewiseboolean}. This comes down to making a consistent choice of atom in the Boolean subalgebras with two atoms, corresponding to the atypical cases for AW*-algebras before.

Returning to C*-algebras, Lemma~\ref{lem:partition} reduces the question of characterizing $\cC(A)$ for a C*-algebra $A$ to finding relationships between $\cC(A)$ and $\cC(C)$ for $C \in \cC(A)$. One prototypical case where we know such a relationship is for the $n$-level system $A=\M_n(\C)$. Namely, inspired by the previous section, there is an action of the unitary group $U(n)$ on $\cC(A)$: if $u \in U(n)$ is some rotation, and $C \in \cC(A)$ is diagonal in some basis, then also the rotation $uCu^*$ is diagonal in the rotated basis and therefore is in $\cC(A)$ again. 
In fact, \emph{any} $C \in \cC(A)$ will be a rotation of an element of $\cC(A)$ that is diagonal in the standard basis. Therefore, we can recognize $\cC(\M_n(\C))$ as a \emph{semidirect product} of $\cC(\mathbb{C}^n)$ and $U(n)$. Such semidirect products can be axiomatized; for details, we refer to~\cite{heunen:cccc}. This can be generalized to C*-algebras $A$ that have a \emph{weakly terminal} commutative C*-subalgebra $D$, in the sense that any $C \in \cC(A)$ allows an injection $C \to D$. This includes all finite-dimensional C*-algebras, as well as algebras of all bounded operators on a Hilbert space.
For example, for the $n$-level system $A=\M_n(\C)$, the matrices that are diagonal in the standard basis form a terminal subalgebra $\C^n$.

However, the mere partially ordered set $\cC(A)$ cannot detect this unitary action. For this we need injections rather than inclusions. Therefore, we now switch to a \emph{category} $\cC_{\rightarrowtail}(A)$ of commutative C*-subalgebras, with \emph{injective $*$-homomorphisms} between them. 
For $A=\M_n(\C)$, these morphisms consist of a rotation in $U(n)$ followed by an inclusion $\C^k \to C^l$ with $k \leq l$.
The following theorem characterizes this category $\cC_{\rightarrowtail}(A)$ up to equivalence. This is the same as characterizing $\cC(A)$ up to \emph{Morita equivalence}, meaning that it determines the topos of contextual sets on $\cC(A)$ discussed in Section~\ref{sec:topos} up to categorical equivalence, rather than determining $\cC(A)$ itself up to equivalence. To~phrase the following theorem, we introduce the monoid $S(X)$ of continuous surjections $X \to X$ on a compact Hausdorff space $X$. In the finite-dimensional case, this is just the symmetric group $S(n)$. Because of our switch from $\cC(A)$ to $\cC_{\rightarrowtail}(A)$, it plays the role of the unitary group we need.

\begin{Theorem}[\cite{heunen:cccc}]\label{thm:semidirect}
  Suppose that a C*-algebra $A$ has a weakly terminal commutative C*-subalgebra $C(X)$.
  A~category is equivalent to $\cC_{\rightarrowtail}(A)$ if and only if it is equivalent to a semidirect product of $\cC(C(X))$ and $S(X)$.
\end{Theorem}

See also~\cite{florifritz:compositiriesgleaves}.

The unitary action can also be used to determine $\cC(A)$ for small $A$ such as $\M_n(\C)$. Combining Lemma~\ref{lem:partition} with Theorem~\ref{thm:semidirect}, we see that $k$-dimensional $C$ in $\cC(\M_n(\C))$ are parametrized by a partition of $n$ into $k$ nonempty parts together with an element of $U(n)$. Two such parameters induce the same subalgebra when the unitary permutes equal-sized parts of the partition. This can be handled neatly in terms of \emph{Young tableaux} and \emph{Grassmannians}, see~\cite{heunenlandsmanspitters:bohrification,fauserraynaudvickers:bundles}. 

Using this concrete parametrization of $\cC(A)$ for $A=\M_n(\C)$, to characterize $\cC(A)$ it would suffice to characterize the unitary group $U(A)$.
Surprisingly, this question is open, even in the finite-dimensional case. All that seems to be known is that, up to isomorphism, $U(1)$ is the unique nondiscrete locally compact Hausdorff group all of whose proper closed subgroups are finite~\cite{morris:unitaries}.
This~characterization does not generalize to finite dimensions higher than one, although closed subgroups have received study in the infinite-dimensional case~\cite{kadison:unitaries}. 
The unitary group $U(n)$ is also, up to isomorphism, the unique irreducible subgroup of $\mathrm{GL}(n)$ the trace of whose elements is bounded~\cite{marcusnewman:unitarygroups}.
It is known that unitary groups of C*-algebras cannot be countably classified~\cite{kerrlupiniphillips:borelautomorphisms}.
Finally, the characterization of $\cC(B(H))$ for Hilbert spaces $H$ could give rise to a description of the category of Hilbert spaces in terms of generators and relations~\cite{heunen:ltwo}.

\section{Generalizations}\label{sec:cqm}

As mentioned in the introduction, the idea to describe quantum structures in terms of their classical substructures applies very generally. This final section discusses to what extent algebraic quantum theory is special, by considering a generalization as an example of another framework. 

Namely, we consider \emph{categorical quantum mechanics}~\cite{heunenvicary:cqm}. This approach formulates quantum theory in terms of the category of Hilbert spaces, and then abstracts away to more general categories with the same structures. Specifically, what is retained is the notion of a \emph{tensor product} to be able to build compound systems, the notion of \emph{entanglement} in the form of objects that form a duality under the tensor product, and the notion of \emph{reversibility} in the sense that every map between Hilbert spaces has an adjoint in the reverse direction. It turns out that these primitives suffice to derive a lot of quantum-mechanical features, such as scalars, the Born rule, no-cloning, quantum teleportation, and complementarity. As a case in point, one can define so-called \emph{Frobenius algebras} in any category with this structure, which is important because of the following proposition.

\begin{Theorem}[\cite{vicary:quantumalgebras,abramskyheunen:hstar}]\label{thm:frobenius}
  Finite-dimensional C*-algebras correspond to Frobenius algebras in the category of Hilbert spaces.
\end{Theorem}

The point is that these notions make sense in \emph{any} category with a tensor product, entanglement, and reversibility. A different example of such a category is that of sets with relations between them. That is, objects are sets $X$, and arrows $X \to Y$ are relations $R \subseteq X \times Y$. For the tensor product, we take the Cartesian product of sets, which makes every object dual to itself and thereby fulfulling the structure of entanglement, and time reversibility is given by taking the opposite relation $R^\dag \subseteq Y \times X$. Two relations $R \subseteq X \times Y$ and $S \subseteq Y \times Z$ compose to $S \circ R = \{(x,z) \mid \exists y \colon (x,y) \in R, (y,z) \in S \}$. We~may regard this as a toy example of \emph{possibilistic quantum theory}: rather than complex matrices, we~now care about entries ranging over $\{0,1\}$. A \emph{groupoid} is a small category, every arrow of which is an isomorphism; they may be considered as a multi-object generalization of groups.

\begin{Theorem}[\cite{heunencontrerascattaneo:groupoids}]\label{thm:groupoids}
  Frobenius algebras in the category of sets and relations correspond to groupoids.
\end{Theorem}

Algebraic quantum theory, as set out in the introduction, makes perfect sense in categories such as sets and relations as well~\cite{coeckeheunenkissinger:channels}. However, in this generality, it is not true that all classical subsystems determine a quantum system at all. The previous theorem provides a counterexample. In commutative groupoids, there can only be arrows $X \to X$, for arrows $g \colon X \to Y$ between different objects cannot commute with their inverse, as $g \circ g^{-1}=1_Y$ and $g^{-1} \circ g = 1_X$. Therefore, any arrow between different objects in a groupoid can never be recovered from any commutative subgroupoid. 

Similarly, quantum logic, as discussed in Section~\ref{sec:topos}, makes perfect sense in this general categorical setting~\cite{heunenjacobs:kernels}. Moreover, it matches neatly with algebraic quantum theory via taking projections~\cite{heunen:complementarity}. However, it is no longer true that commutative subalgebras correspond to Boolean sublattices. Again, a counterexample can be found using Theorem~\ref{thm:groupoids}~\cite{coeckeheunenkissinger:logic}.

One could object that commutativity might be too narrow a notion of classicality. However, consider broadcastability instead: classical information can be broadcast, but quantum information cannot. More precisely, a Frobenius algebra $A$ is \emph{broadcastable} when there exists a completely positive map $A \to A \otimes A$ such that both partial traces are the identity $A \to A$. Again, this makes perfect sense in general categories. It turns out that the broadcastable objects in the category of sets and relations are the groupoids that are totally disconnected, in the sense that there are no arrows $g \colon X \to Y$ between different objects~\cite{heunenvicary:cqm}. So even with this more liberal operational notion of classicality, classical subsystems do not determine a quantum system.

This breaks a well-known information-theoretic characterization of quantum theory, that is phrased in terms of C*-algebras~\cite{cliftonbubhalvorson,heunenkissinger:cbh}. Hence there is something about (algebraic) quantum theory beyond the categorical properties of having tensor products, entanglement, and reversibility, that~underwrites Bohr's doctrine of classical concepts. 
It relates to characterizing unitary groups, as~discussed in Section~\ref{sec:characterization}. 
We close this overview by raising the interesting interpretational question of just what this defining property is.

\nocite{*}
\bibliographystyle{plain}
\bibliography{classicalfaces}

\end{document}